\documentclass[sigconf]{acmart}

\usepackage{booktabs}

\AtBeginDocument{%
  \providecommand\BibTeX{{%
    \normalfont B\kern-0.5em{\scshape i\kern-0.25em b}\kern-0.8em\TeX}}}
\usepackage{svg}
\usepackage{tcolorbox}
\usepackage{url}
\usepackage{amsmath}
\usepackage{graphicx}
\usepackage{colortbl}
\usepackage{subcaption}
\usepackage{algorithm}
\usepackage{algpseudocode}

\copyrightyear{2024}
\acmYear{2024}
\setcopyright{acmlicensed}
\acmBooktitle{28th International Conference on Evaluation and Assessment in Software Engineering (EASE 2024), June 18--21, 2024, Salerno, Italy}
\acmDOI{10.1145/3661167.3661225}
\acmISBN{979-8-4007-1701-7/24/06}

\acmConference[EASE 2024]{28th International Conference on Evaluation and Assessment in Software Engineering}{June 18--21, 2024}{Salerno, Italy}

\begin{document}

\newcommand{\marcio}[1]{{\color{red} [#1]}}
\newcommand{\elvys}[1]{{\color{blue} #1}}
\newcommand{\manoel}[1]{{\color{purple} #1}}
\newcommand{\davi}[1]{{\color{orange} #1}}
\newcommand{\NumberOfSurveyedProfessionals}{15}
\newcommand{\fmeasure}{$83.70\%$}
\newcommand{\recall}{$80.85\%$}
\newcommand{\precision}{$86.75\%$}
\newcommand{\surveyApprovalMean}{$91.43\%$}
\newcommand{\TotalNumberOfTestsWithinFiles}{$973$}
\newcommand{\TotalSmellOccurrences}{$8,386$}
\newcommand{\SampleOfSmellOccurrences}{$264$}
 
\title{A Catalog of Transformations to Remove Smells From\\Natural Language Tests}

\author{Manoel Aranda}
\orcid{0000-0001-9540-1605}
\affiliation{%
  \institution{Federal University of Alagoas}
  \city{Maceió}
  \country{Brazil}
}
\email{mpat@ic.ufal.br}

\author{Naelson Oliveira}
\orcid{0000-0002-3232-7257}
\affiliation{%
  \institution{Federal University of Alagoas}
  \city{Maceió}
  \country{Brazil}
}
\email{naelson@ic.ufal.br}

\author{Elvys Soares}
\orcid{0000-0001-7593-0147}
\affiliation{
  \department{Federal Institute of Alagoas} 
  \city{Maceió}
  \country{Brazil}
}
\email{elvys.soares@ifal.edu.br}

\author{M\'{a}rcio Ribeiro}
\orcid{0000-0002-4293-4261}
\affiliation{
  \department{Federal University of Alagoas}
  \city{Maceió}
  \country{Brazil}
}
\email{marcio@ic.ufal.br}

\author{Davi Romão}
\orcid{0009-0000-8173-6068}
\affiliation{%
  \institution{Federal University of Alagoas}
  \city{Maceió}
  \country{Brazil}
}
\email{dsr@ic.ufal.br}

\author{Ullyanne Patriota}
\orcid{0009-0000-5034-8214}
\affiliation{%
  \institution{Federal University of Alagoas}
  \city{Maceió}
  \country{Brazil}
}
\email{ufjp@ic.ufal.br}

\author{Rohit Gheyi}
\orcid{0000-0002-5562-4449}
\affiliation{
  \department{Federal Univ. of Campina Grande}
  \city{Campina Grande}
  \country{Brazil}
}
\email{rohit@dsc.ufcg.edu.br}

\author{Emerson Souza}
\orcid{0009-0008-0322-2262}
\affiliation{
  \department{Federal University of Pernambuco} 
  \city{Recife}
  \country{Brazil}
}
\email{epss@cin.ufpe.br}

\author{Ivan Machado}
\orcid{0000-0001-9027-2293}
\affiliation{
  \department{Federal University of Bahia}
  \city{Salvador}
  \country{Brazil}
}
\email{ivan.machado@ufba.br}

\renewcommand{\shortauthors}{Aranda, et al.}

\begin{abstract}
Test smells can pose difficulties during testing activities, such as poor maintainability, non-deterministic behavior, and incomplete verification. Existing research has extensively addressed test smells in automated software tests but little attention has been given to smells in natural language tests. 
While some research has identified and catalogued such smells, there is a lack of systematic approaches for their removal. Consequently, there is also a lack of tools to automatically identify and remove natural language test smells.
This paper introduces a catalog of transformations designed to remove seven natural language test smells and a companion tool implemented using Natural Language Processing (NLP) techniques. Our work aims to enhance the quality and reliability of natural language tests during software development.
The research employs a two-fold empirical strategy to evaluate its contributions. First, a survey involving \NumberOfSurveyedProfessionals\ software testing professionals assesses the acceptance and usefulness of the catalog's transformations. Second, an empirical study evaluates our tool to remove natural language test smells by analyzing a sample of real-practice tests from the Ubuntu OS.
The results indicate that software testing professionals find the transformations valuable. Additionally, the automated tool demonstrates a good level of precision, as evidenced by a F-Measure rate of \fmeasure.
\end{abstract}



\maketitle

\section{Introduction}
\label{sec:intro}

Test smells are indications of bad decisions when designing or implementing tests. Such decisions may introduce potential problems in testing activities~\cite{van2001refactoring}. Examples of these problems are poor maintanability (\textit{e.g.}, duplication in test code~\cite{hauptmann2012clones}), non-deterministic scenarios~\cite{meszaros2007xunit}, and missing verifications~\cite{spadini2020thresholds} due to code branches. In this context, the literature is particularly vast when considering test smells in automated software tests. Previous studies have been introduced to provide catalogues of test smells~\cite{almeida2015catalog,garousi2018smells}, to count and study their occurrences~\cite{bavota2015test,soares2023tse,soares2020refactoring}, and to provide tools capable of identifying and removing them automatically using code transformations~\cite{soares2020refactoring,soares2023tse,luanaTestSmells,van2001refactoring, luanna-refactoring-mining}.

\begin{figure}[htb]
    \centering
    \includegraphics[width=.90\columnwidth]{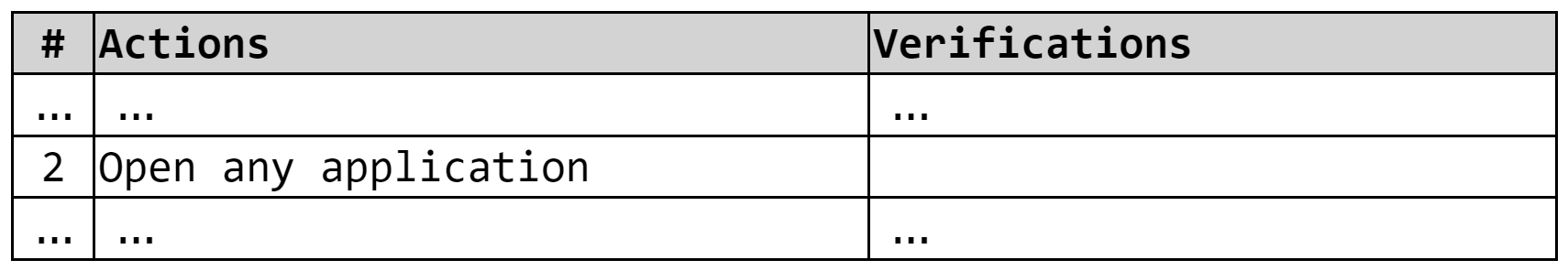}
    \caption{Natural Language Test Smell: Ambiguous Test.}
    \label{fig:introduction:example}
    \Description[Example of Natural Language Test Smell]{A table with three columns, #, Actions and Verifications. On the second line, there is an action "Open any application", and there is no verification. All other lines are non-important}
\end{figure}

Unfortunately, the same does not apply when considering test smells in manual tests, \textit{i.e.}, natural language tests. These tests are written in natural language and also suffer from problems introduced by smells~\cite{hauptmann2013hunting,soares2023manual}. Figure~\ref{fig:introduction:example} illustrates part of a natural language test extracted from the Ubuntu Operating System (Ubuntu OS)~\cite{launchpadUbuntuManual}. The test contains the \textit{Ambiguous Test} smell~\cite{hauptmann2013hunting}, which indicates an ``\textit{under-specified test that leaves room for interpretation.}'' In Step 2, the required action is to ``\textit{Open any application.}'' However, it is not specified which application should be opened. This ambiguity can lead to different test results based on the application selected by the tester. Hence, identifying and removing this smell might bring better quality standards for the tests.

As mentioned, differently from test smells in automated software tests, test smells in natural language are still poorly explored. Hauptmann \textit{et al.}~\cite{hauptmann2013hunting} presented seven natural language test smells and after ten years Soares \textit{et al.}~\cite{soares2023manual} complemented the list with six additional smells. Both studies introduced rules to identify the test smells. However, we still miss a catalog of transformations to remove natural language test smells. In addition, we miss tool support to make this whole process automatic, \textit{i.e.}, a tool capable of removing natural language test smells automatically.

To minimize these problems, in this paper we introduce a catalog of transformations to remove natural language test smells presented in recent research~\cite{soares2023manual}. Our catalog contains seven transformations capable of removing the following natural language test smells: \textit{Unverified Action}, \textit{Misplaced Precondition}, \textit{Misplaced Action}, \textit{Misplaced Verification}, \textit{Eager Action}, \textit{Ambiguous Test}, and \textit{Conditional Test}. For each smell, our catalog contains details with respect to the \textit{addressed test smell}, the mechanics to apply the \textit{transformation} in the test in order to remove such smell, and an \textit{example}. The catalog works considering a left-hand side test, \textit{i.e.}, the test that contains the test smell; and a right-hand side test, \textit{i.e.}, the test after the transformation without the smell. We also contribute to a tool based on Natural Language Processing (NLP) techniques. The tool is capable of identifying the test smells and removing them automatically. In case the test contains more than one smell, the tool works sequentially, tackling one smell at a time.

To evaluate the catalog and the tool, we conduct a two-fold empirical strategy. First, we conduct a survey to answer the research question \textbf{RQ$_{1}$:}\textit{``How software testing professionals perceive and evaluate the transformations of our catalog?''} To answer it, we recruited \NumberOfSurveyedProfessionals\ software testing professionals from a large smartphone manufacturer. The name of the company is omitted due to non-disclosure agreements. Answering \textbf{RQ$_{1}$} is important to better understand whether the software testing professionals find our transformations useful to improve the quality of the tests. Second, we conduct an empirical study to evaluate our tool. Here we focus on the research question \textbf{RQ$_{2}$:}\textit{``How precise is our tool in the task of removing natural language test smells?''} To conduct this study and answer \textbf{RQ$_{2}$}, we consider real-practice natural language tests from the Ubuntu OS. The tests consist of not only software functionalities, but also interactions with the hardware. After executing our tool against \TotalNumberOfTestsWithinFiles\ natural language tests, we identified \TotalSmellOccurrences\ occurrences of the seven smells we focus on in this paper. In this scenario, to evaluate the precision of our tool in removing the smells, we would need to manually analyze more than eight thousand transformations, exceeding our capabilities. Thus, to make our manual analysis feasible, we use the Cochran's Sample Size Formula~\cite{kotrlik2001organizational} and analyze a sample of \SampleOfSmellOccurrences\ randomly selected smell occurrences.

The results of our survey indicate an average acceptance of \surveyApprovalMean\ among the software testing professionals we recruited. Regarding the tool, our manual analysis showed that it achieved \fmeasure\ rate of F-Measure.

In summary, this paper provides the following contributions:

\begin{itemize}
    
    \item A catalog of transformations to remove seven test smells from natural language tests (Section~\ref{sec:catalog});

    \item A survey with software testing professionals to validate the catalog (Section~\ref{sec:catalogevaluation});

    \item The development of a tool that makes our transformations automatic (Section~\ref{sec:tool});

    \item An empirical study to check the precision of our tool using real-practice natural language tests from the Ubuntu OS (Section~\ref{sec:tool-validation}).
    
\end{itemize}

A replication package with all results is available at Figshare \cite{ease2024Replication}. Our tool is available at our companion website \cite{ease2024Tool}.
\section{Natural Language Test Smells}
\label{sec:motivating}

Previous works showed that test smells also exist in natural language tests \cite{hauptmann2013hunting,soares2023manual}. Besides the \textit{Ambiguous Test} smell we discussed in Section~\ref{sec:intro}, we now show three additional test smells in natural language tests. We also discuss why they could be harmful during testing activities. All examples of natural language tests in this paper are extracted from the Ubuntu OS manual test repository~\cite{launchpadUbuntuManual}.

\subsection{Unverified Action}
\label{sec:motivating-unverified-action}

\begin{figure}[ht]
    \centering
    \includegraphics[width=.9\columnwidth]{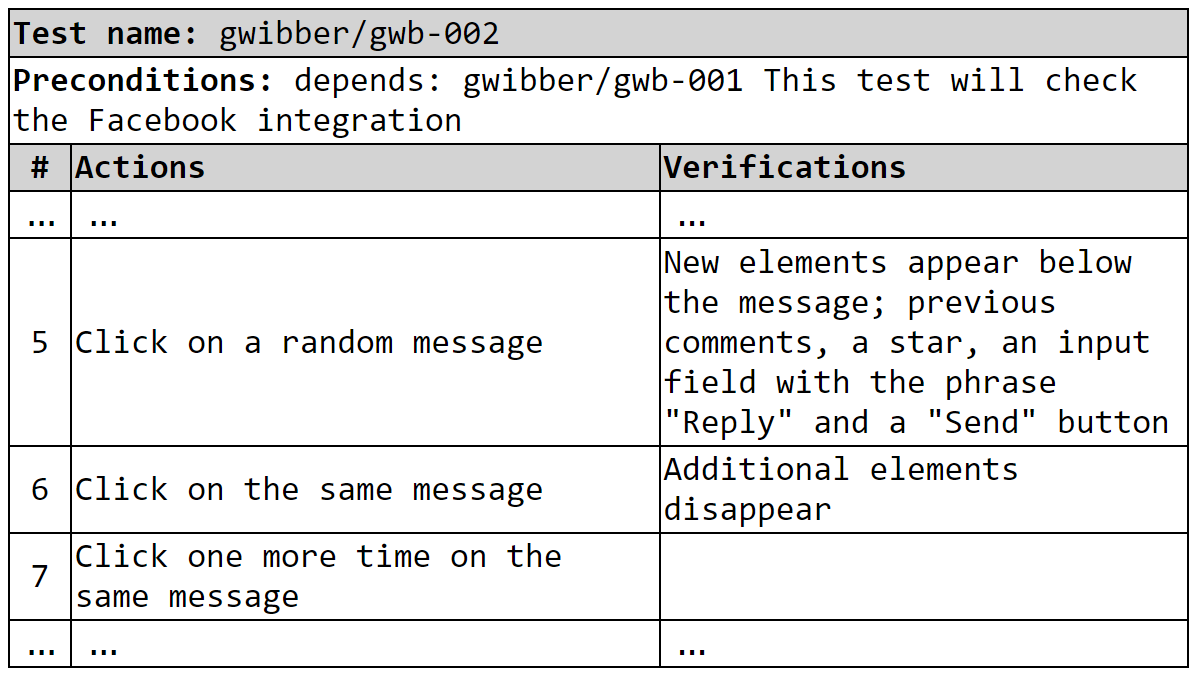}
    \caption{Unverified Action - Example}
    \label{fig:motivating:unverified action}
    \Description[Example of Natural Language Test Smell]{Illustrates part of a natural language test extracted from the Ubuntu Operating System (UbuntuOS). The test contains the Ambiguous Test smell [13], which indicates an “under-specified test that leaves room for interpretation.” In Step 2, the required action is to “Open any application.” However, it is not specified which application should be opened.}
\end{figure}

This smell occurs when there is no verification for a given action. When compared to automated tests, this smell is similar to the \textit{Assertionless} smell, which is defined by the absence of assertions~\cite{aljedaaniAssertionless}. Figure~\ref{fig:motivating:unverified action} presents the \textit{Unverified Action} smell. Step~7 lacks a verification for the ``\textit{Click one more time on the same message}'' action. 

Without proper verification, tests may not effectively validate whether the application is behaving as expected. Regarding the example, the tester may experience confusion regarding the outcomes of the clicking action. Questions such as ``\textit{What should happen when the message is clicked one more time?}'' and ``\textit{Do additional elements disappear? And if they do not, does the test fail?}'' may arise, leaving the tester uncertain on how to proceed. Addressing these uncertainties is crucial to maintain clarity and ensure effective test execution.

\subsection{Eager Action}
\label{sec:motivating-eager-action}

\begin{figure}[h]
    \centering
    \includegraphics[width=.9\columnwidth]{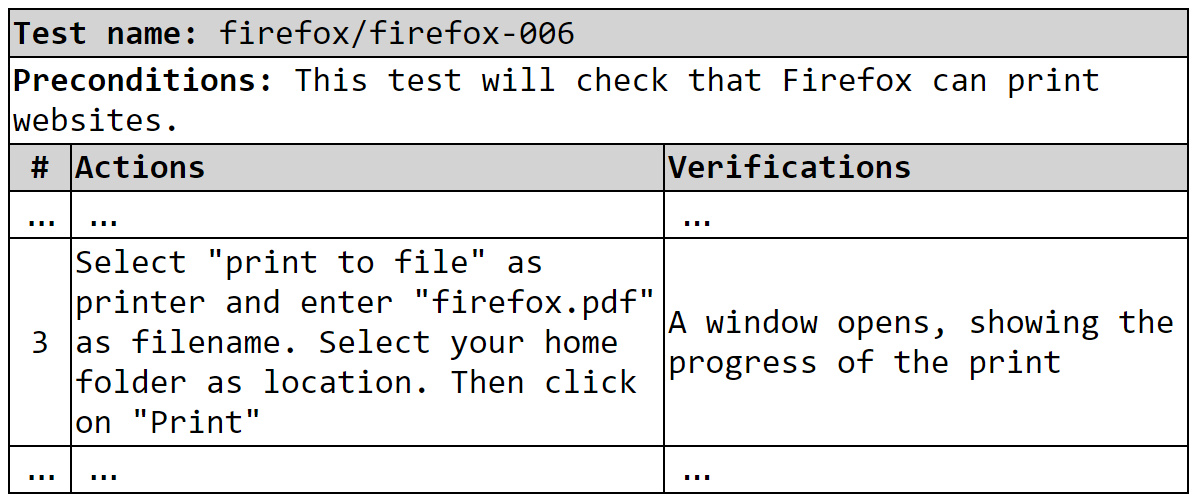}
    \caption{Eager Action - Example}
    \label{fig:motivating:eager action}
    \Description[Eager Action - Example]{This figure depicts an example of the Eager Action smell.}
\end{figure}

This smell happens when a single step groups multiple actions. Figure~\ref{fig:motivating:eager action} depicts an example of the \emph{Eager Action} smell.

The \textit{Eager Action} test smell can introduce problems such as non-isolated tests and difficulties in debugging. In our example, we have four actions, represented by the verbs ``\textit{select}'', ``\textit{enter}'', ``\textit{select}'', and ``\textit{click}.'' Since we have one verification for four actions, in the event of a failure, it may be unclear which action caused it.

\subsection{Conditional Test}
\label{sec:motivating-conditional-test}

\begin{figure}[h]
    \centering
    \includegraphics[width=.9\columnwidth]{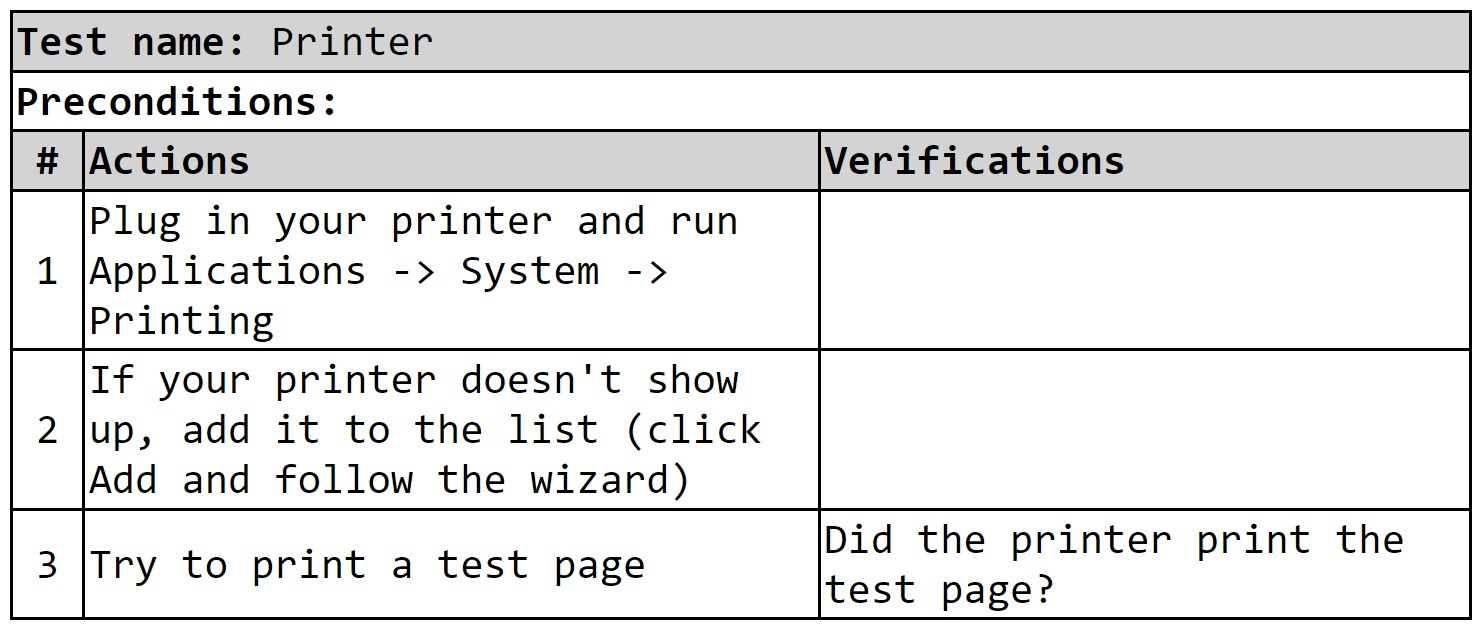}
    \caption{Conditional Test - Example}
    \label{fig:motivating:conditional}
\end{figure}

This smell is defined by a conditional clause appearing in a step description~\cite{hauptmann2013hunting}. Figure~\ref{fig:motivating:conditional} shows an example.

The conditional introduces uncertainty and variability to the test, as the tester may not know how to proceed or what to expect. Notice that the action in Step 2 contains a conditional clause ``\textit{If your printer doesn't show up, add it to the list (click Add and follow the wizard)}.'' If the printer is already showing up, the tester might need assistance deciding whether to skip Step 2 and proceed to Step 3 or stop the test execution altogether because there is no mention of the printer being on the list. This situation can reduce the test's determinism, maintainability, coverage, and reliability~\cite{soares2023tse}.
\section{Removing Natural Language Test Smells}
\label{sec:catalog}

This section introduces a catalog of proposals to remove test smells in natural language tests. Concerning our terminology, according to Fowler \emph{et al.}~\cite{fowler1997refactoring}, refactorings preserve the observable behavior of a code. On the other hand, van Deursen \emph{et al.}~\cite{van2001refactoring} define test refactorings as changes in test code that do not add or remove test cases. Since our proposals do not strictly align with Fowler's and van Deursen's definitions because they may alter the test behavior and even introduce new test cases, we call them \textit{transformations}.

Our catalog contains transformations in terms of a left-hand side and a right-hand side~\cite{soares2023tse}. The left-hand side holds the problematic test (\textit{i.e.}, the test with test smells). The right-hand side presents the test transformed, \textit{i.e.}, the test without test smells, after the transformations. To apply a transformation, the catalog uses semantic pattern matching. In other words, a natural language test that matches the left-hand side of a transformation is converted to the right-hand side.

\subsection{Natural Language Test Template}
\label{sec:catalog:template}

To better explain our transformations, we first need to introduce a template to represent natural language tests. Here, we define a natural language test as $T = (P, S_1, S_2, S_3, \ldots, S_i, \ldots, S_n)$, where:

\begin{itemize}
    
    \item $P$ is a boolean expression, representing the preconditions to execute $T$;
    
    \item $S_i$ is a tuple representing the i-th step of the test. This tuple consists of an ordered list of actions $A_i$ and an ordered list of verifications $V_i$, defined as $S_i = (A_i, V_i)$;

    \item The elements of the $A_i$ list are denoted by $[a_{i1}, a_{i2}, \ldots, a_{in}]$;

    \item The elements of the $V_i$ list are denoted by $[v_{i1}, v_{i2}, \ldots, v_{in}]$.

\end{itemize}

Thus, given the above definitions, the following equations are interchangeable:

\begin{equation}
    S_i = (A_i, V_i)
    \label{eq:si-as-tuple}
\end{equation}
\begin{equation}
    S_i = ([a_{i1}, a_{i2}, \ldots, a_{in}], [v_{i1}, v_{i2}, \ldots, v_{in}])
    \label{eq:si-with-elements}
\end{equation}

Considering the test illustrated in Figure~\ref{fig:motivating:eager action}, we have:

\begin{itemize}

    \item $P = $ ``\textit{This test will check that Firefox can print websites}'';
    
    \item $A_3 = $ [``\textit{Select `print to file' as printer}'', ``\textit{enter `firefox.pdf' as filename}'', ``\textit{Select your home folder as location}'', ``\textit{Then click on `Print'}''];
    
    \item $V_3 = $ [``\textit{A window opens, showing the progress of the print}''].

\end{itemize}

\begin{figure}[h]
    \centering
    \includegraphics[width=.9\columnwidth]{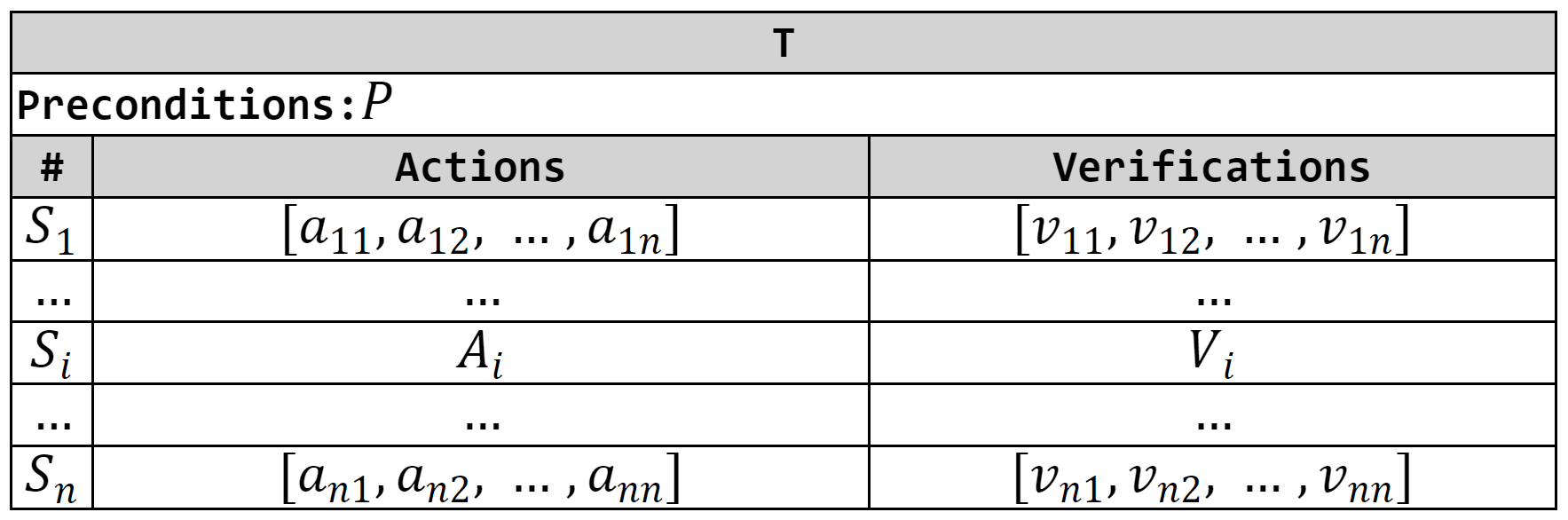}
    \caption{Natural Language Test Template}
    \label{fig:catalog:template}
\end{figure}

Figure~\ref{fig:catalog:template} presents $T$ and all its elements in terms of a table. Notice that in Step $S_i$ we use the form presented in Equation~\ref{eq:si-as-tuple}. When considering Steps $S_1$ and $S_n$ we use the form presented in Equation~\ref{eq:si-with-elements}.

By using this template and particularly the Step $S_i$, we can discuss how to avoid natural language test smells. For example: to avoid the \textit{Ambiguous Test} smell, the sentences in $A_i$ and in $V_i$ should not be ambiguous; To avoid the \textit{Eager Action} smell in $S_i$, the cardinality of the list $A_i$ should be $|A_i| = 1$, \emph{i.e.}, there should be only one action instead of several ones in the i-th step; To avoid the \textit{Unverified Action} smell in $S_i$, the cardinality of the list $V_i$ should be $|V_i| \geq 1$, \emph{i.e.}, there is at least one verification in $V_i$.

When presenting our transformations, we also consider the infix $++$ operator, which is used for list concatenation. It takes two lists as operands and combines them to form a new list. Equation \ref{eq:catalog:operator} shows a simple usage of the $++$ operator.

\begin{equation} \label{eq:catalog:operator}
    \begin{split}
        S_i = & [a_{i1}, a_{i2}, \ldots, a_{in}] ++ ~[a_k] \\
        S_i = & [a_{i1}, a_{i2}, \ldots, a_{in}, a_k]
    \end{split}
\end{equation}

\subsection{A Catalog of Transformations}

We now present our catalog having each transformation in terms of (i) the smell that the transformation addresses, (ii) the mechanics to apply the transformation (left-hand side and right-hand side), (iii) implications---in case the transformation removes a smell but adds another, or the transformation removes more than one smell ---, and (iv) an example of the transformation in a natural language test extracted from the Ubuntu OS.

In this paper, we propose transformations for seven natural language test smells, namely \textit{Unverified Action}, \textit{Misplaced Precondition}, \textit{Misplaced Action}, \textit{Misplaced Verification}, \textit{Eager Action}, \textit{Ambiguous Test}, and \textit{Conditional Test}. We focus on these smells because they may introduce difficulties in testing activities and have been studied by the literature~\cite{hauptmann2013hunting, soares2023manual}.

\subsubsection{Fill Verification} 
\label{sec:fill-verification-transformation}

\paragraph{Addressed Smell} The \textit{Unverified Action}. We discussed this smell in Section~\ref{sec:motivating-unverified-action}.

\begin{figure}[htb]
  \begin{subfigure}[b]{\columnwidth}
  \centering
    \includegraphics[width=.9\columnwidth]{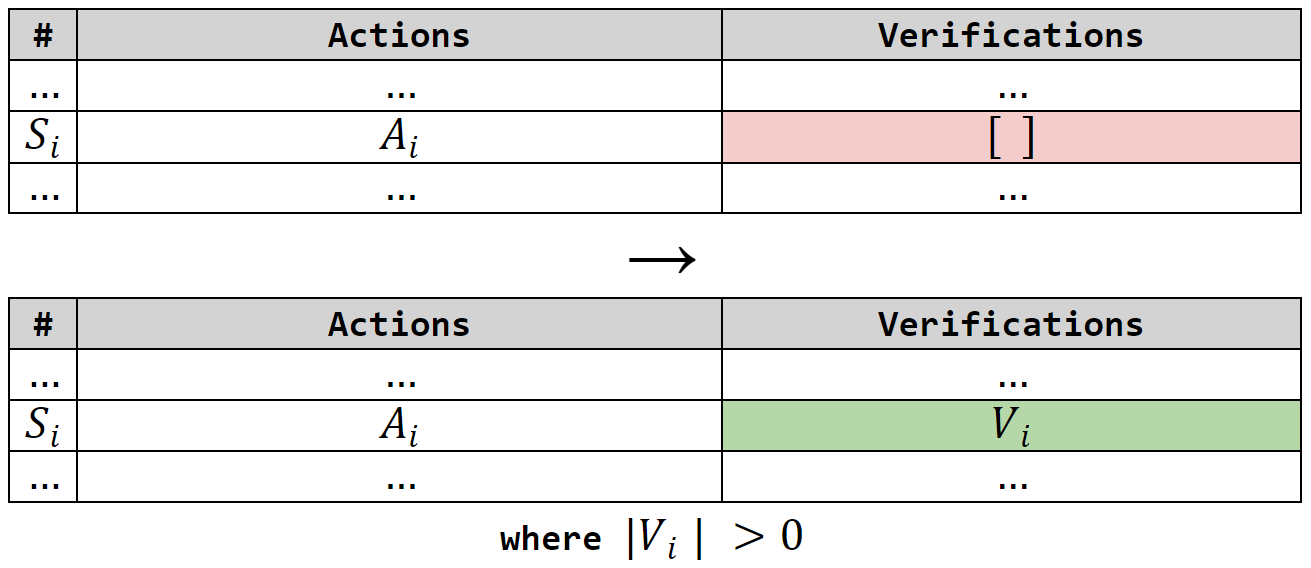}
    \caption{Transformation}
    \label{fig:catalog:fill verification}
  \end{subfigure}
  \\
  \begin{subfigure}[b]{\columnwidth}
  \centering
    \includegraphics[width=.9\columnwidth]{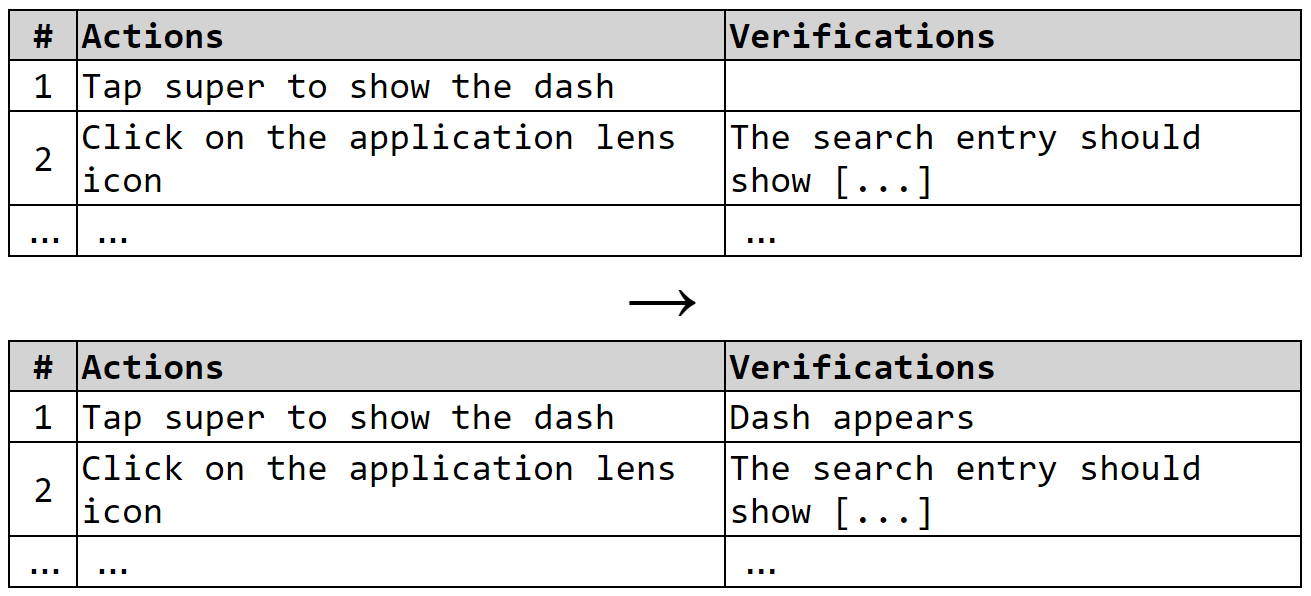}
    \caption{Example}    
    \label{fig:catalog:fill verification:example bonitinho}
  \end{subfigure}
  \caption{Fill Verification}
  \label{fig:catalog:fillVerificationAB}
\end{figure}

\textit{Formalization}. Figure \ref{fig:catalog:fill verification} presents our transformation to remove the \textit{Unverified Action} smell. Notice that, at the left-hand side, the verifications list $V_i$ is empty. To remove the smell, we can implement two strategies. The first and simplest one is to just warn the tester that $V_i$ is empty and then add a sort of ``\texttt{FILL\_VERIFICATION}'' flag in the \texttt{Verifications} field. The second strategy needs to deal with more complex algorithms. As $A_i$ and $V_i$ should be somehow linked by the same context, we can use NLP techniques to infer the verification from the action. Then, we add the inferred verification sentence in $V_i$.

\textit{Example}. Figure~\ref{fig:catalog:fill verification:example bonitinho} shows an example of the \textit{Fill Verification} transformation. In this example, we add the ``\textit{Dash appears}'' verification based on the text contained in the actions list $A_1$.

\subsubsection{Extract Precondition}
\label{sec:catalog:extract-precondition}

\paragraph{Addressed Smell} The \textit{Misplaced Precondition}, which appears when a precondition is placed as the first action of the test. This situation might bring difficulties in test correctness. For example, the tester might report a test failure. However, what actually happened was that a precondition to execute the test was not met.

\begin{figure}[htb]
  \begin{subfigure}[b]{\columnwidth}
  \centering
    \includegraphics[width=.9\columnwidth]{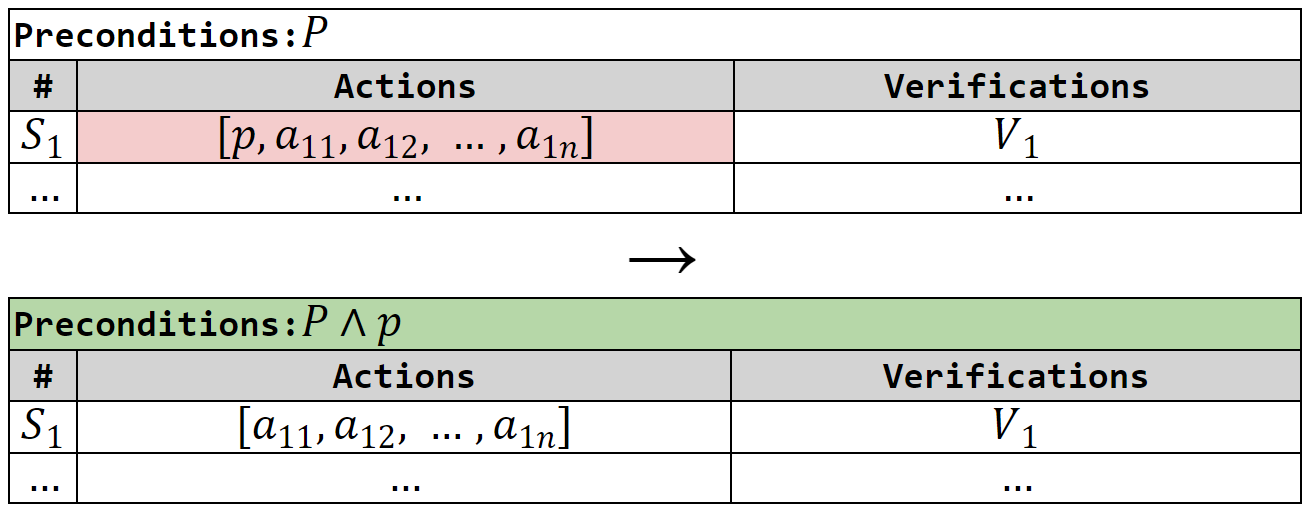}
    \caption{Transformation}
    \label{fig:catalog:extract precondition}
  \end{subfigure}
  \\
  \begin{subfigure}[b]{\columnwidth}
  \centering
    \includegraphics[width=.9\columnwidth]{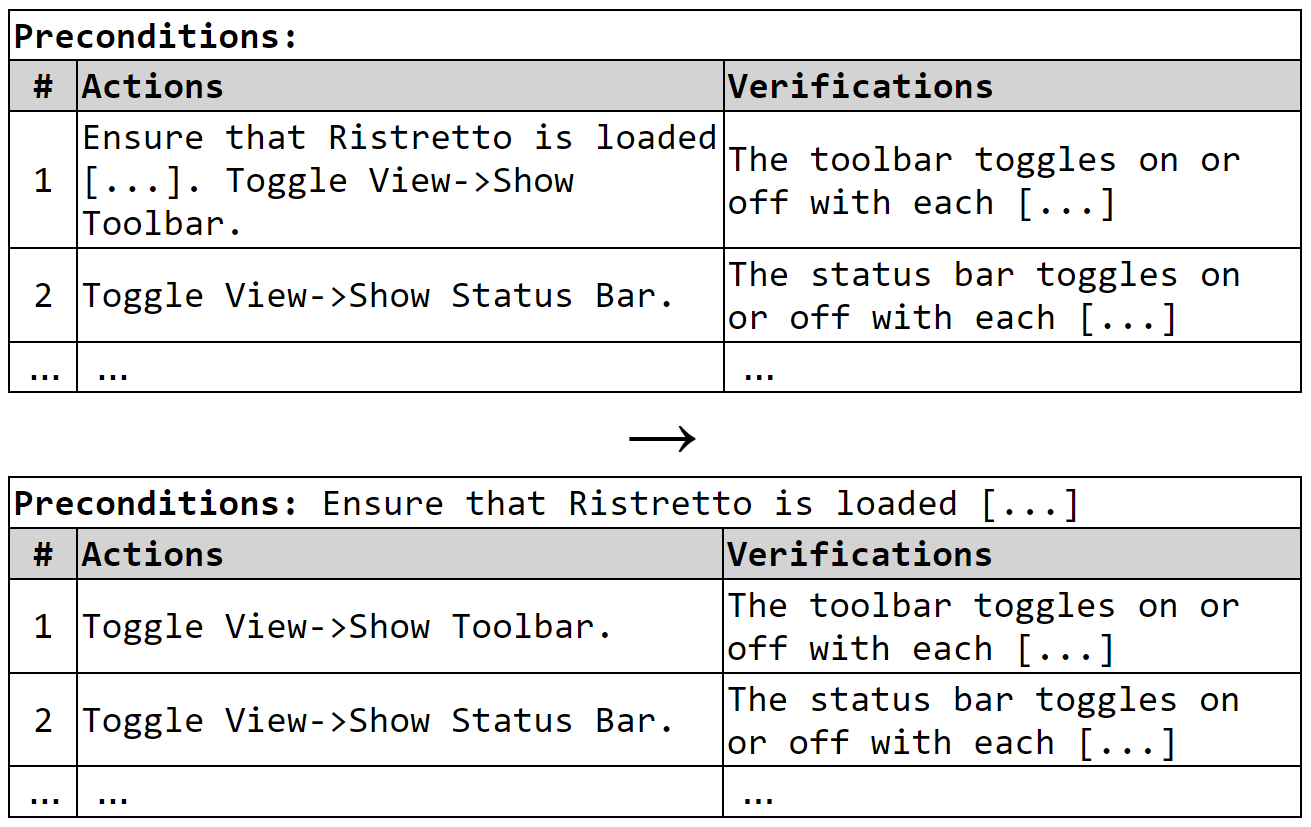}
    \caption{Example}    
    \label{fig:catalog:extract precondition:example}
  \end{subfigure}
  \caption{Extract Precondition}
  \label{fig:catalog:ExtractPreconditionAB}
\end{figure}

\textit{Formalization}. Figure~\ref{fig:catalog:extract precondition} shows a precondition $p$ as the first element of $A_1$. We deal with the smell by removing $p$ from the actions list $A_1$ and placing it in the boolean expression $P$ using the conjunction operator, yielding $P \land p$.

\textit{Example}. Figure~\ref{fig:catalog:extract precondition:example} presents an example of applying our \textit{Extract Precondition} transformation. Notice that the precondition ``\textit{Ensure that Ristretto is loaded [...]}'' was removed from the actions list and added to the \texttt{Preconditions} field.

\subsubsection{Extract Action} 
\label{sec:catalog:extract-action}

\paragraph{Addressed Smell} The \textit{Misplaced Action}, which happens when there is an action in the verifications list. Such confusion can result in inconsistent execution and potentially misguide the interpretation of results.

\begin{figure}[htb]
  \begin{subfigure}[b]{\columnwidth}
    \centering
    \includegraphics[width=.9\columnwidth]{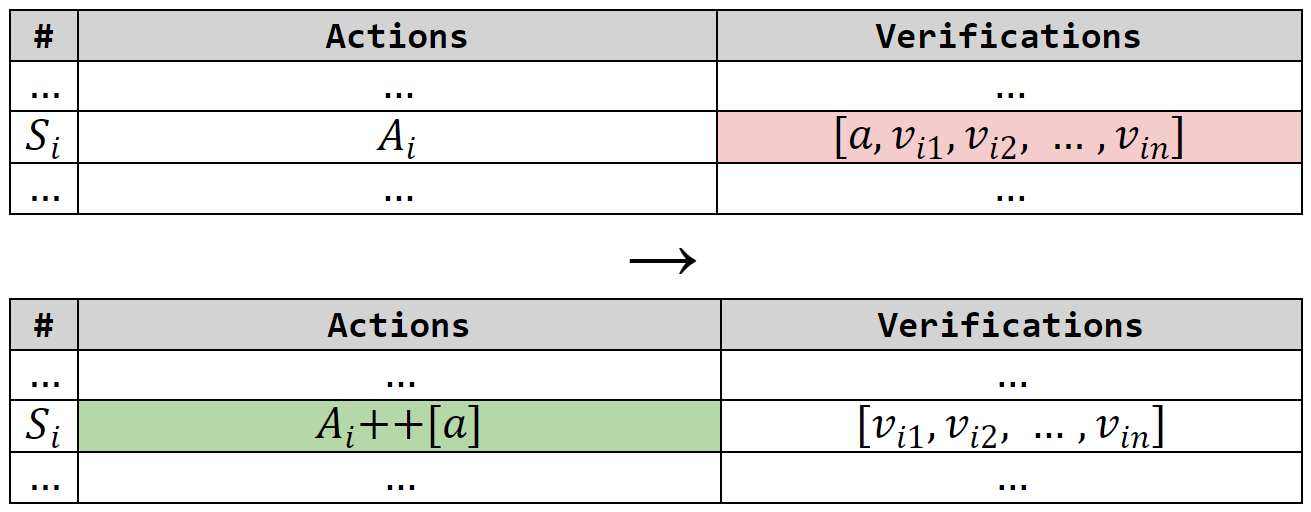}
    \caption{Transformation}
    \label{fig:catalog:extract action}
  \end{subfigure}
  \\
  \begin{subfigure}[b]{\columnwidth}
    \centering
    \includegraphics[width=.9\columnwidth]{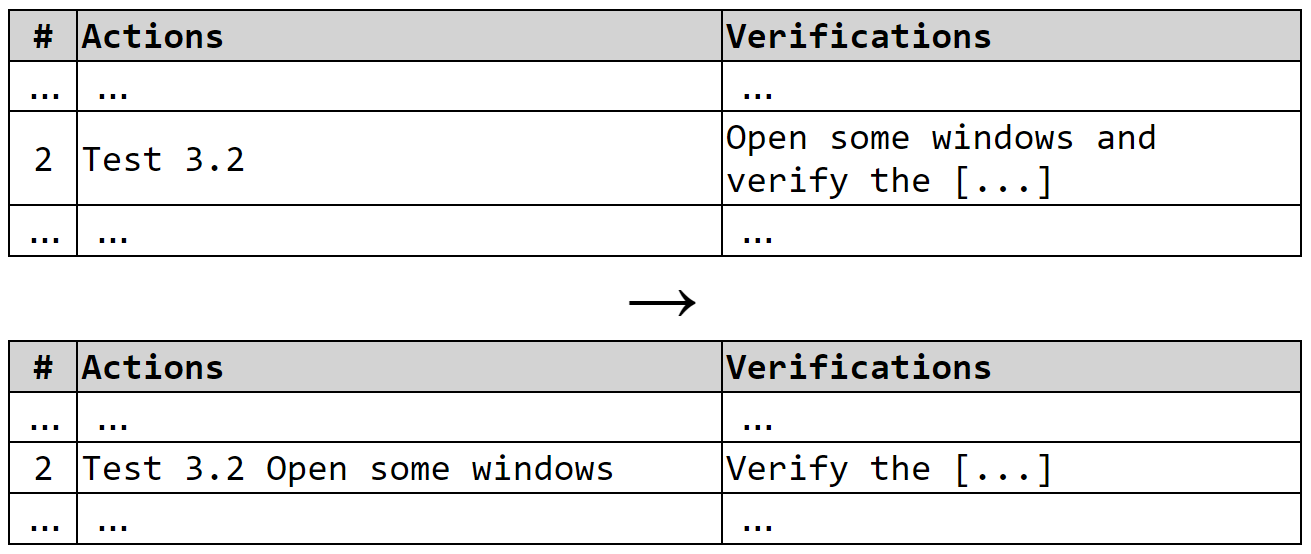}
    \caption{Example}    
    \label{fig:catalog:extract action:example}
  \end{subfigure}
  \caption{Extract Action}
  \label{fig:catalog:extractactionAB}
\end{figure}

\textit{Formalization}. Figure~\ref{fig:catalog:extract action} illustrates the action element $a$ in the verifications list of Step $S_i$. To remove the smell, all we need to do is to remove $a$ from the verifications list $V_i$ and add it to the actions list $A_i$. To perform the adding task, we use the $++$ operator.

\textit{Implications}. The \textit{Extract Action} transformation might introduce the \textit{Eager Action} smell in case there already exists an action in $A_i$. In this sense, to remove the \textit{Eager Action} smell, we propose the \textit{Separate Actions} transformation. We introduce this transformation in Section~\ref{sec:separate-actions-transformation}.

\textit{Example}. Figure~\ref{fig:catalog:extract action:example} presents a test with the action ``\textit{Open some windows}'' located in the verifications list. After executing our transformation, such an action is placed in the actions list and removed from the verifications list.

\subsubsection{Extract Verification}

\paragraph{Addressed Smell} The \textit{Misplaced Verification}, which occurs when there is a verification in the actions list.

\begin{figure}[htb]
  \begin{subfigure}[b]{\columnwidth}
    \centering
    \includegraphics[width=.9\columnwidth]{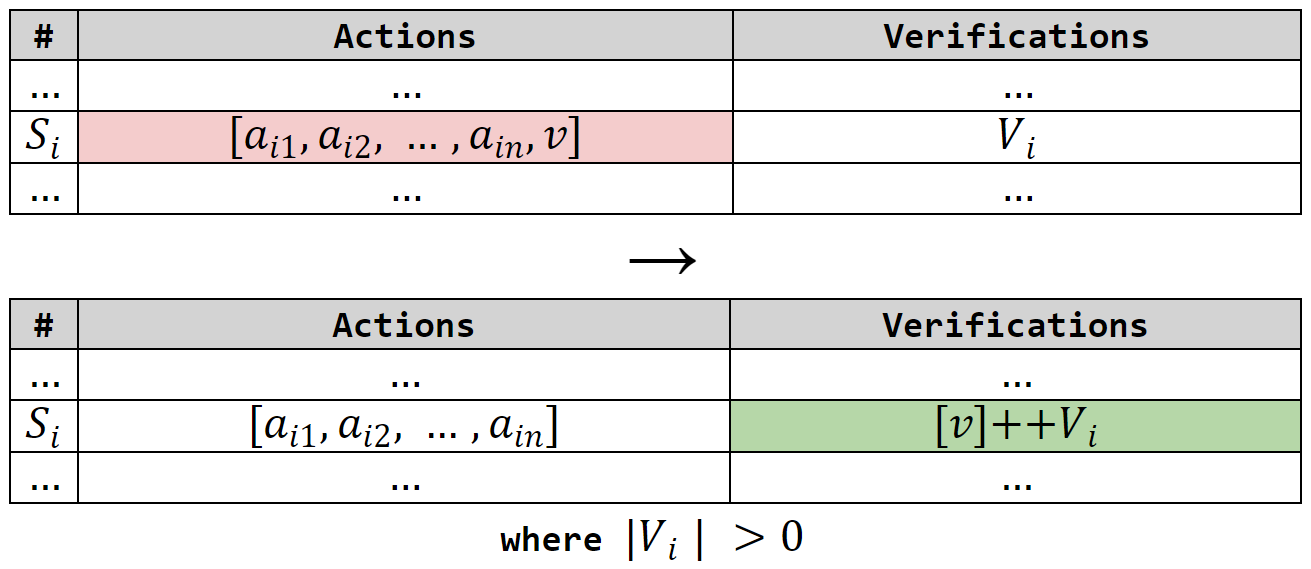}
    \caption{Transformation}
    \label{fig:catalog:extract verification}
  \end{subfigure}
  \\
  \begin{subfigure}[b]{\columnwidth}
    \centering
    \includegraphics[width=.9\columnwidth]{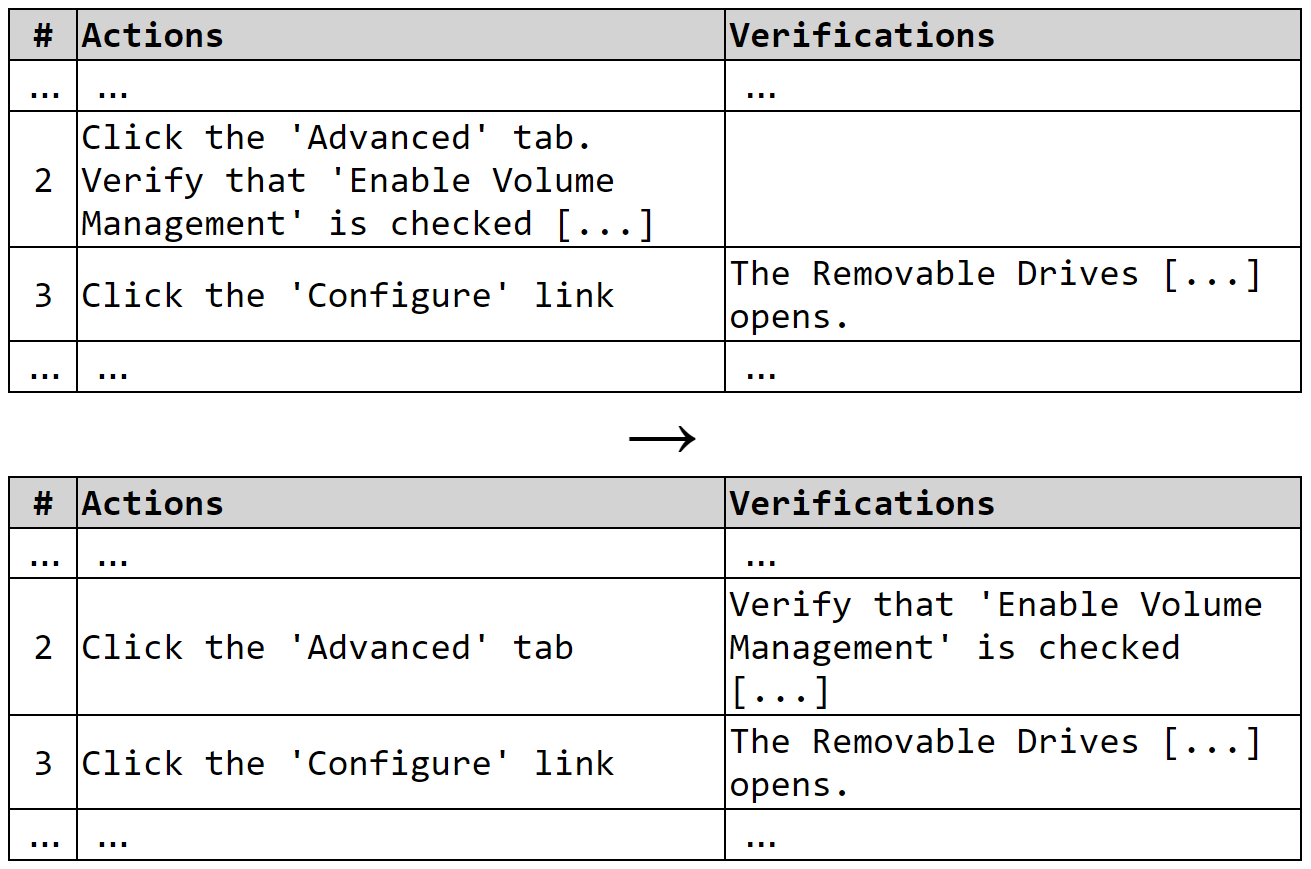}
    \caption{Example}
    \label{fig:catalog:extract verification:example}
  \end{subfigure}
  \caption{Extract Verification}
  \label{fig:catalog:extractverificationAB}
\end{figure}

\textit{Formalization}. Figure~\ref{fig:catalog:extract verification} illustrates a verification $v$ in $A_i$ list. To address the smell, we remove $v$ from the actions list $A_i$ and add it to the verifications list $V_i$. In this transformation, we rely on the $++$ operator. Notice that we add $v$ at the beginning of $V_i$. We opt for this because $v$ should be the first verification to be checked right after executing action $a_{in}$.

\textit{Implications}. Applying this transformation might remove not only the \textit{Misplaced Verification}, but also the \textit{Unverified Action} smell. This happens in case $|V_i| = 0$ and we move $v$ from $A_i$ to $V_i$, making $|V_i| = 1$.

\textit{Example}. Figure~\ref{fig:catalog:extract verification:example} illustrates the verification ``\textit{Verify that `Enable Volume Management' is checked [...]}'' in the actions list. After our transformation, we place such a verification in the verifications list. With this transformation, notice that we remove the \textit{Misplaced Verification} and the \textit{Unverified Action} smells.

\subsubsection{Separate Actions}
\label{sec:separate-actions-transformation}

\paragraph{Addressed Smell} The \textit{Eager Action}, discussed in Section~\ref{sec:motivating-eager-action}.

\begin{figure}[htb]
    \begin{subfigure}[b]{\columnwidth}
    \centering
    \includegraphics[width=.91\columnwidth]{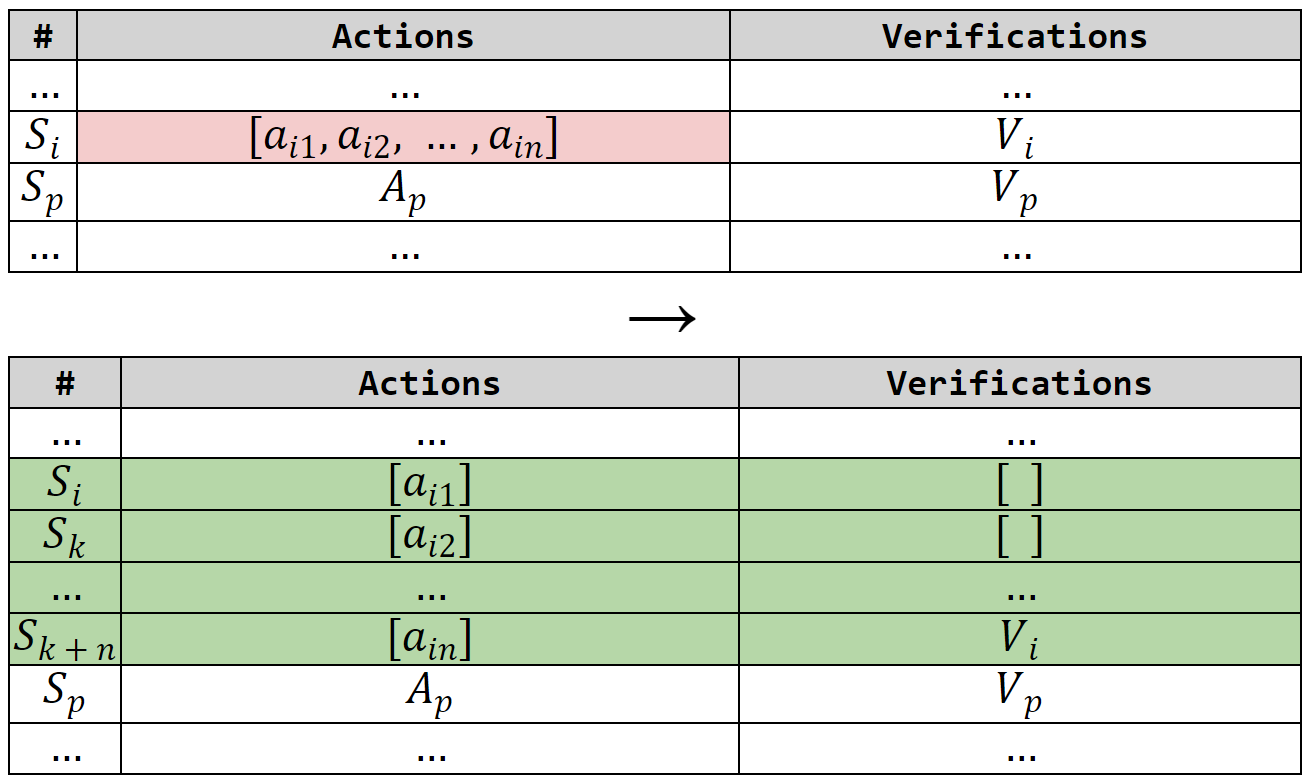}
    \caption{Transformation}
    \label{fig:catalog:separate actions}
    \end{subfigure}
    \\
    \begin{subfigure}[b]{\columnwidth}
    \centering
    \includegraphics[width=.91\columnwidth]{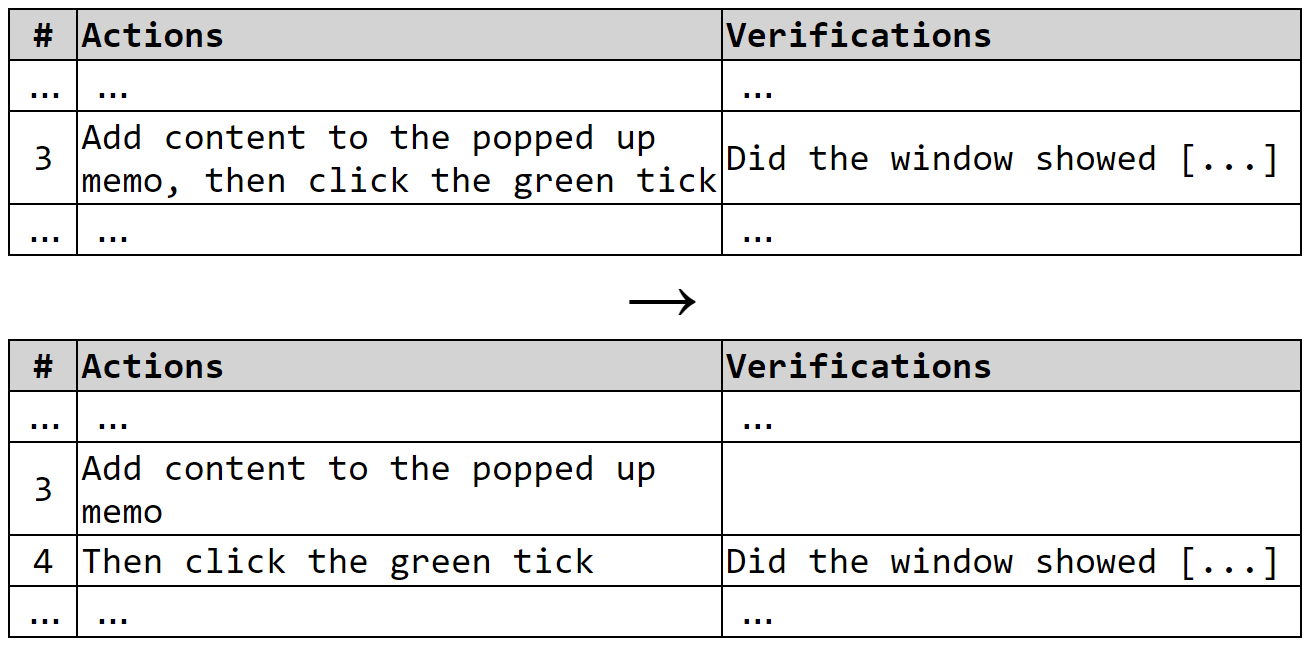}
    \caption{Example}
    \label{fig:catalog:separate actions:example bonito}
    \end{subfigure}
    \caption{Separate Actions}
    \label{fig:catalog:separateActionsAB}
  \end{figure}

\textit{Formalization}. Figure~\ref{fig:catalog:separate actions} illustrates that $A_i$ contains $n$ elements, characterizing the \textit{Eager Action} smell: $|A_i| > 1$. To remove the smell, we define a new step for each element of the $A_i$ list. Additionally, in our transformation, we associate the verifications list $V_i$ with the step created for the last action originally in $A_i$, \textit{i.e.}, the $a_{in}$ action. We opt for this because we consider that the tester will check the verifications list $V_i$ only after executing the last action.

\textit{Implications}. The \textit{Separate Actions} transformation may lead to the \textit{Unverified Action} smell. After the transformation, we have empty verifications lists for the Steps from $S_i$ to $S_{k + n - 1}$. To deal with the \textit{Unverified Action} smell, we introduce the \textit{Fill Verification} transformation (Section~\ref{sec:fill-verification-transformation}).

\textit{Example}. Figure~\ref{fig:catalog:separate actions:example bonito} shows Step~3 with two actions: ``\textit{Add content to the popped up memo}'' and ``\textit{Then click the green tick}.'' According to our transformation, they should be split in two different steps. Moreover, the verification ``\textit{Did the window showed [...]}'' has been associated with the last action, \textit{i.e.}, ``\textit{Then click the green tick}.''

\subsubsection{Extract Ambiguity}
\label{sec:catalog:catalog:removeAmbiguities}

\paragraph{Addressed Smell} The \textit{Ambiguous Test} smell. Ambiguities and under-specified activities may lead to problems during the test execution, such as different outcomes.
 
\begin{figure}[htb]
    \begin{subfigure}[b]{\columnwidth}
    \centering
    \includegraphics[width=.9\columnwidth]{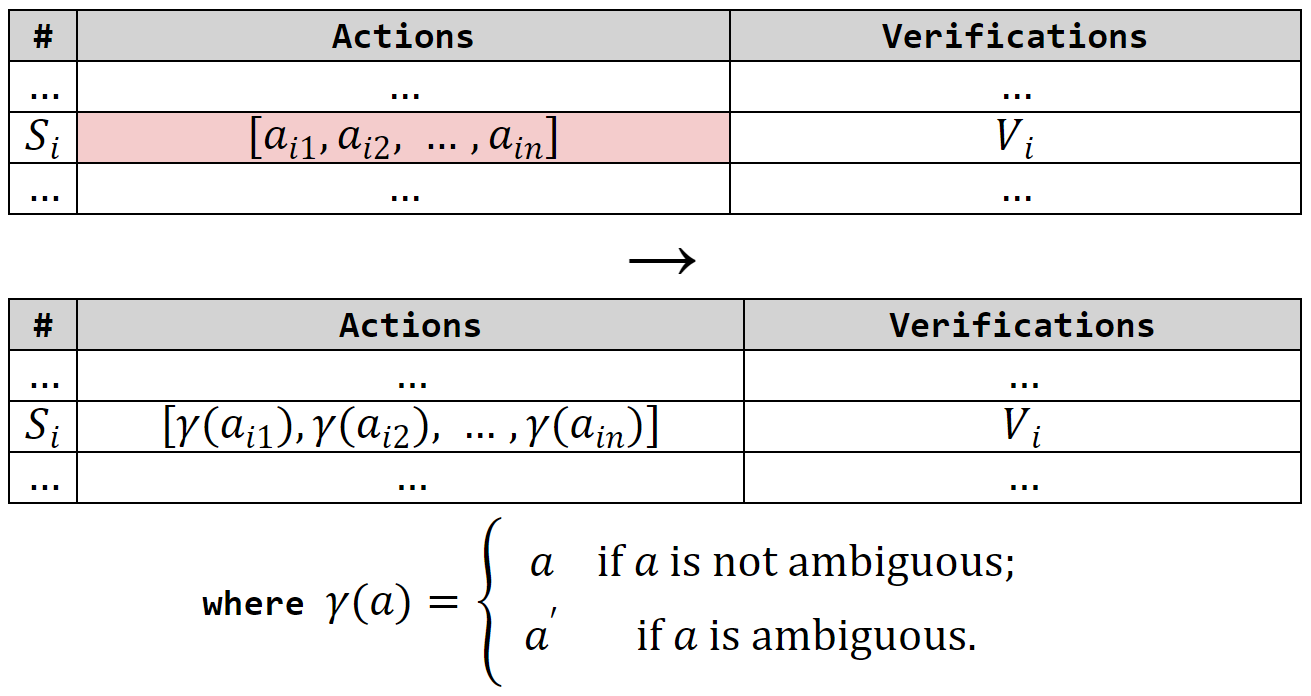}
    \caption{Transformation}
    \label{fig:catalog:remove ambiguities}
    \end{subfigure}
    \\
    \begin{subfigure}[b]{\columnwidth}
    \centering
    \includegraphics[width=.9\columnwidth]{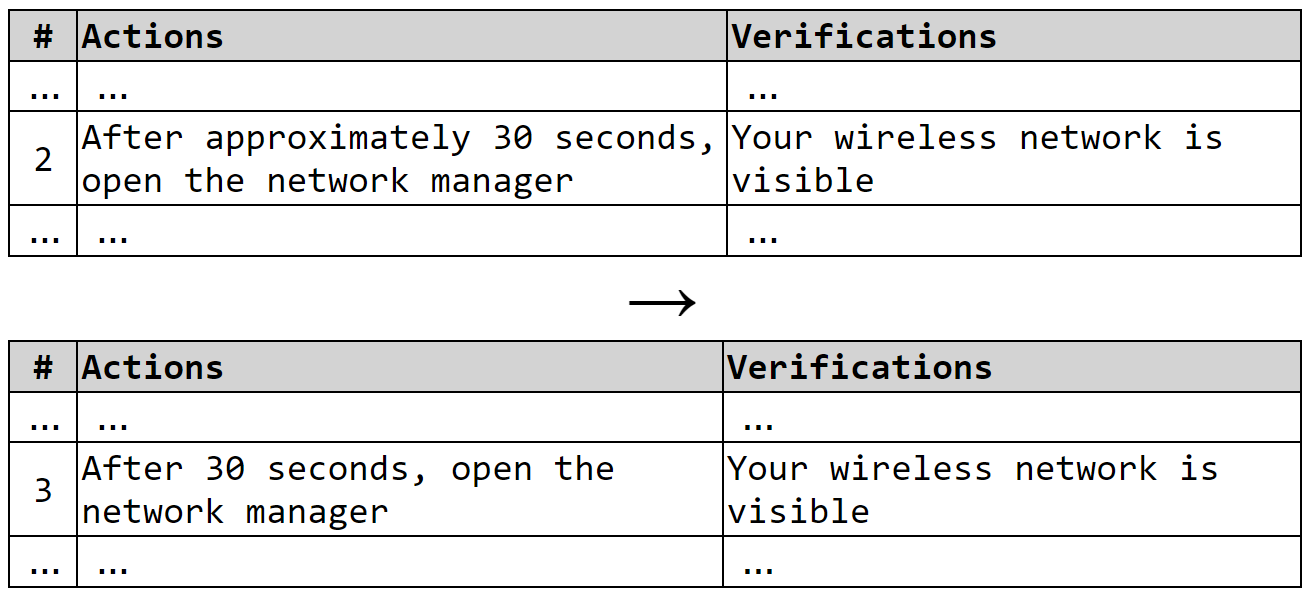}
    \caption{Example}
    \label{fig:catalog:remove ambiguities:example}
    \end{subfigure}
    \caption{Extract Ambiguity}
    \label{fig:catalog:extractAmbiguityAB}
  \end{figure}

\textit{Formalization}. Figure~\ref{fig:catalog:remove ambiguities} shows the actions list $A_i$ with $n$ elements. Any of these elements might have ambiguous sentences, such as ``\textit{quickly}'' or ``\textit{accurately}.'' To remove the ambiguities, we propose the use of a $\gamma$ function that receives an action sentence as input and returns the same action sentence in case it is not ambiguous; and a modified action sentence in case the original one is ambiguous. We then apply $\gamma$ to all elements of $A_i$. Although our transformation to remove the \textit{Ambiguous Test} smell focused on the actions list, we can use the same rationale on the verifications list.

\textit{Example}. Figure~\ref{fig:catalog:remove ambiguities:example} illustrates an action with the adverb of manner ``\textit{approximately}.'' Here the tester could ask: ``\textit{Is 25 seconds enough to make the wireless network visible? If I wait 25 seconds and the wireless network does not appear, did the test fail?}''

\subsubsection{Extract Conditional}

\paragraph{Addressed Smell} The \textit{Conditional Test} smell, detailed in Section~\ref{sec:motivating-conditional-test}. To deal with this smell, we create two different tests~\cite{jannik-jss}.

\begin{figure}[htb]
    \begin{subfigure}[b]{\columnwidth}
    \centering
    \includegraphics[width=.9\columnwidth]{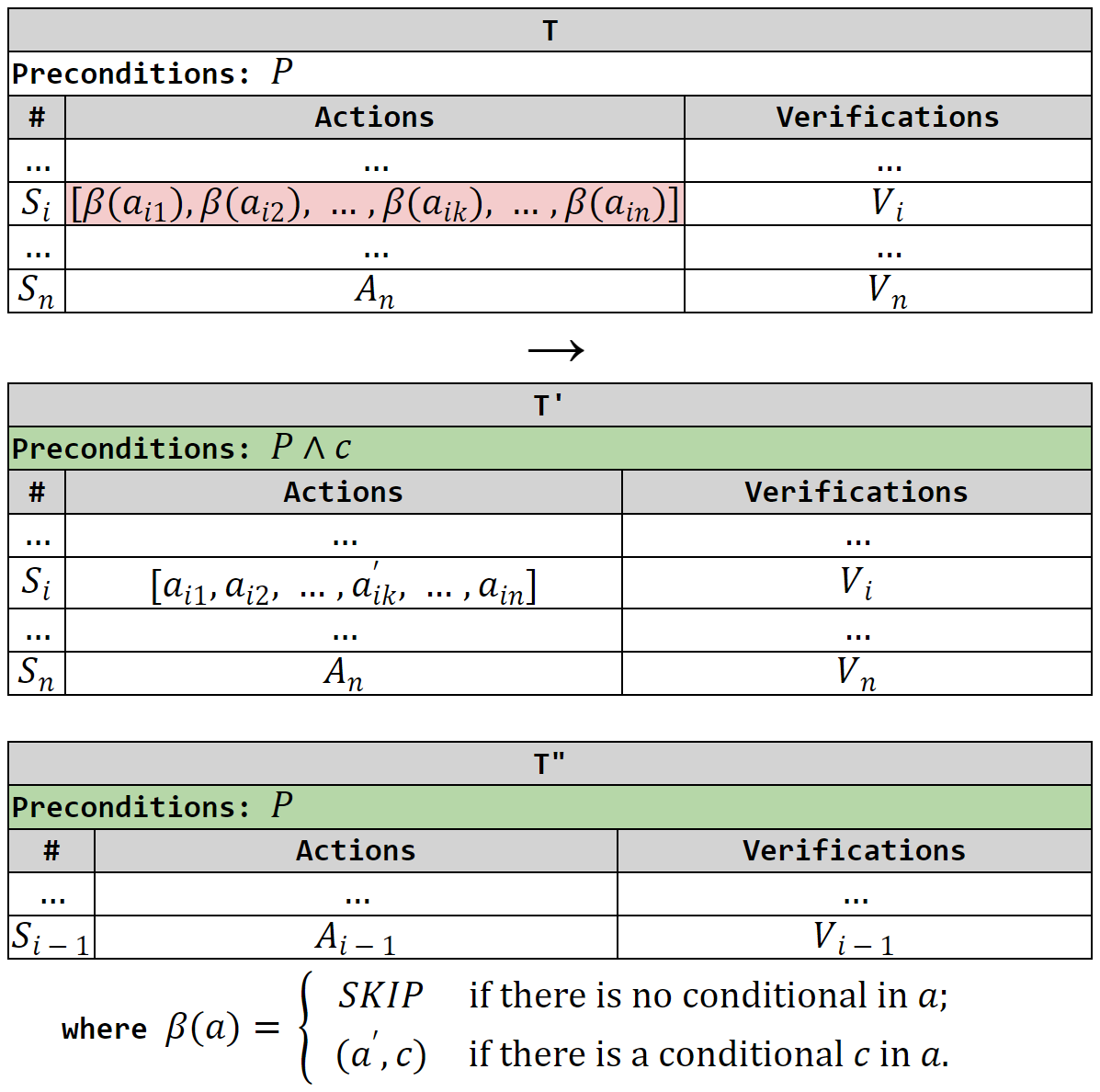}
    \caption{Transformation}
    \label{fig:catalog:remove conditionals}
    \end{subfigure}
    \\
    \begin{subfigure}[b]{\columnwidth}
    \centering
    \includegraphics[width=.9\columnwidth]{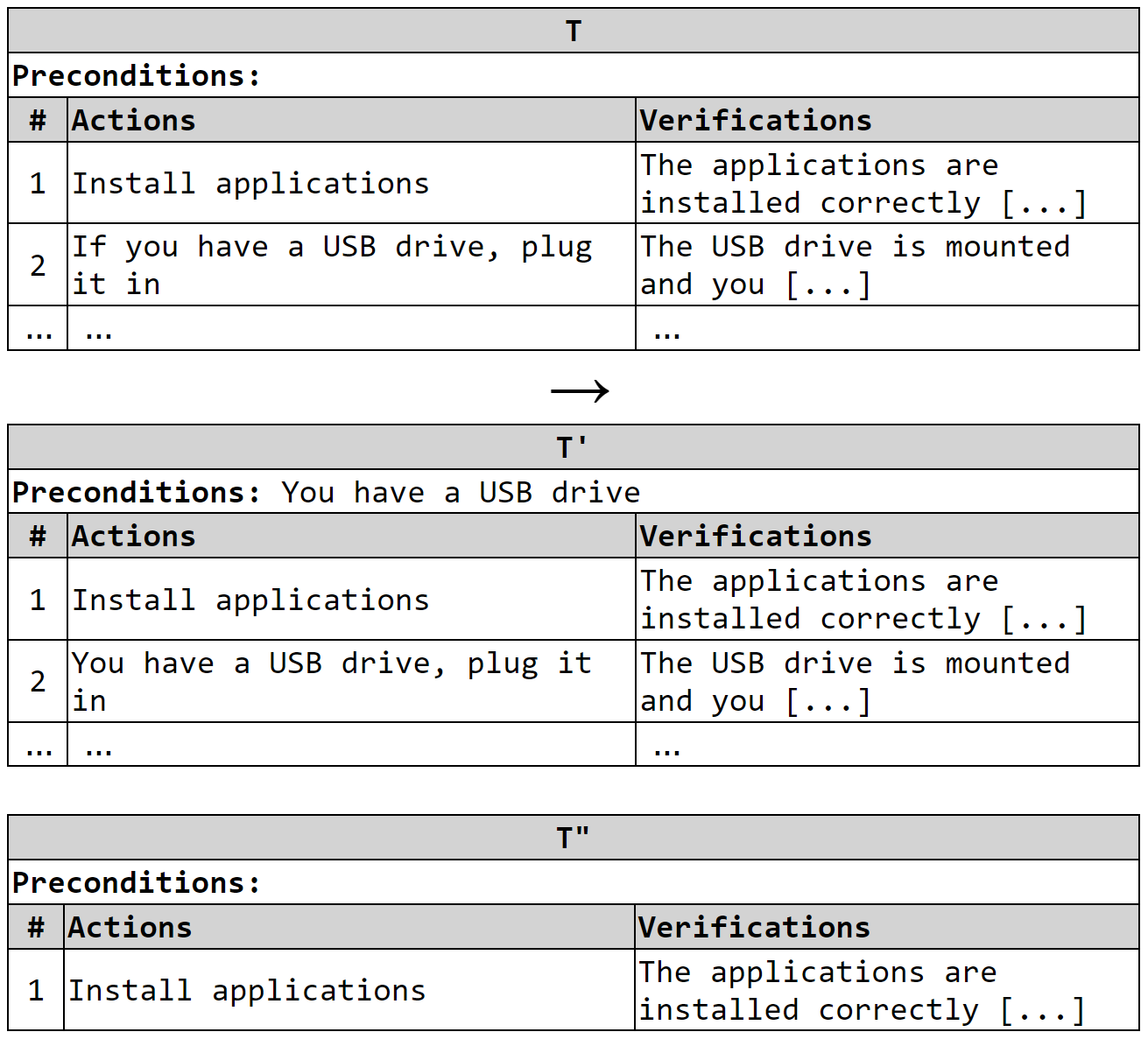}
    \caption{Example}
    \label{fig:catalog:remove conditionals:example}
    \end{subfigure}
    \caption{Extract Conditional}
    \label{fig:catalog:extractConditionalAB}
  \end{figure}

\textit{Formalization}. Figure~\ref{fig:catalog:remove conditionals} illustrates our transformation. To do so, we rely on two functions. The first one, $\beta$, receives an action sentence $a$ as input. In case $a$ has no conditional, $\beta$ simply skips. In case $a$ has a conditional $c$, $\beta$ returns a map $(a',c)$, where $a'$ is the action element without the conditional and $c$ is the removed conditional. In Figure~\ref{fig:catalog:remove conditionals}, $\beta$ identified a conditional $c$ in action $a_{ik}$ and removed it, transforming $a_{ik}$ into $a_{ik}'$. 
Because now we have the conditional $c$, it is time to create two tests. We consider another function $\theta$ that receives the original test $T$ and the conditional $c$. $\theta$ yields two tests: $T'$ considering $c$ as true; and $T''$ considering $c$ as false. Because the conditional $c$ is necessary to execute $T'$, we place it in the boolean expression $P$ using the conjunction operator, yielding $P \land c$. As to $T''$, we exclude the step that originally contained the conditional and the following steps. Notice that we need to repeat this process to all actions in all steps of the original test.

\textit{Example}. Figure~\ref{fig:catalog:remove conditionals:example} presents an example of the \textit{Conditional Test} smell removal. Once we identify the conditional ``\textit{If you have a USB drive},'' we create a test considering it in the precondition and another test without the step that originally had the conditional and the subsequent steps.

  \subsubsection{Summary}

Illustrated in Table~\ref{tab:tool:transformations-smells} is the list of our proposed transformations and their addressed smells.

\begin{table}[htb]
\caption{Transformations and addressed smells}
\label{tab:tool:transformations-smells}
\small
\begin{tabular}{@{}rll@{}}
\toprule
\textbf{No.} & \textbf{Transformation} & \textbf{Addressed Smell} \\                                \midrule
1              & Extract Conditional      & Conditional Test                                     \\
2              & Extract Action          & Misplaced Action                                      \\
3              & Separate Actions        & Eager Action                                          \\
4              & Extract Verification    & Misplaced Verification    \\
5              & Extract Ambiguity        & Ambiguous Test                                       \\
6              & Extract Precondition    & Misplaced Precondition                                \\
7              & Fill Verification       & Unverified Action                                     \\ \bottomrule
\end{tabular}%
\end{table}
\section{Catalog Evaluation}
\label{sec:catalogevaluation}

This section presents our evaluation, conducted as an online survey with software testing professionals. Here we aim to answer the research question \textbf{RQ$_{1}$: ``How do software testing professionals perceive and evaluate the transformations of our catalog?''}

\subsection{Planning}

The goal of this study is to evaluate the effectiveness of our transformations. To do so, we assess the opinions of software testing professionals about the transformations through an online survey. By analyzing ``before'' and ``after'' test transformation snippets---originally taken from the Ubuntu OS manual tests and transformed according to our proposals---along with the participants’ comments on their answers, we would be able to validate whether the respondents were aware of any benefits.

Each survey question had a test smell definition, problem, identification steps, original and transformed samples, the query \textit{``Do you agree that, in the example below, the identified problem was addressed  according to the definition?''}, a Likert scale ranging from \textit{``I strongly agree''} to \textit{``I strongly disagree''}, and an optional comments field. Finally, we recruited participants from a large smartphone manufacturer, which received invitations by email.

\subsection{Results}

\begin{figure*}[ht]
  \centering
  \includegraphics[scale=0.25]{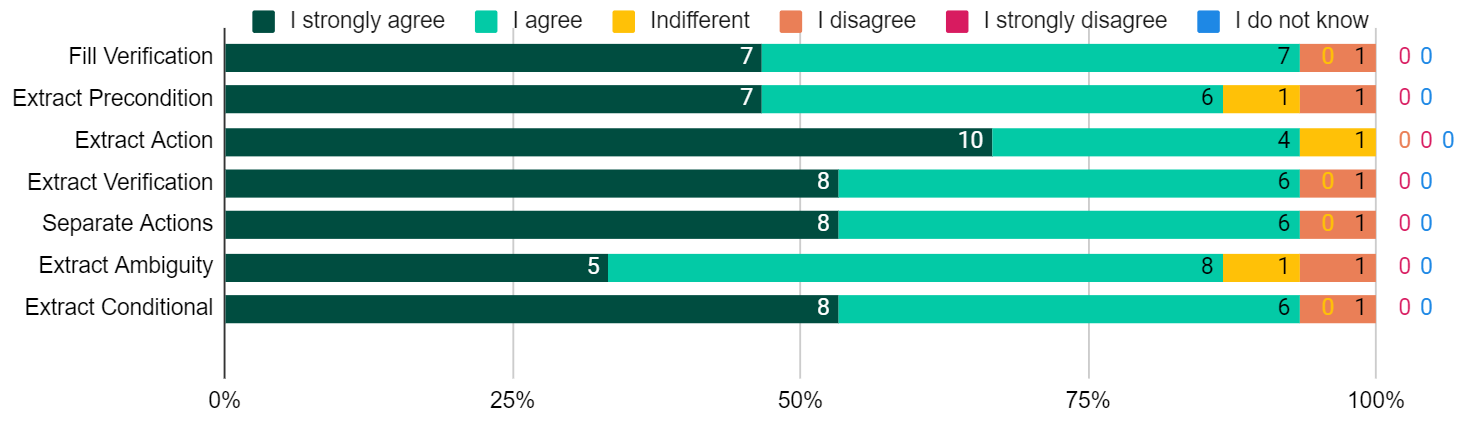}
  \caption{Online Survey Results}
  \label{fig:catalogEvaluation:surveyResults}
\end{figure*}

We performed the survey in November 2023 -- January 2024, achieving \NumberOfSurveyedProfessionals\ responses. Concerning the demographics, 71\% of the participants defined their primary work area as the industry (over academia) and their average experience with software testing was 4,1 years. One participant works in Portugal and 14 in Brazil. Figure~\ref{fig:catalogEvaluation:surveyResults} details the obtained results concerning the participants' opinions about the transformation samples.

\subsection{Discussion}

Regarding the proposed transformations, the opinion of the experienced test professionals served as a validation that obtained a high average acceptance rate of \surveyApprovalMean. There were no disagreements to the \textit{Extract Action} transformation, that was accepted by 93.3\% of the respondents (one respondent was indifferent). Four transformations reached 93.3\% acceptance rate (14 respondents) and 6.7\% of rejection rate (one respondent): \textit{Fill Verification}, \textit{Separate Actions}, \textit{Extract Verification}, and \textit{Extract Conditional}. Our transformations with the lowest acceptance rate were the \textit{Extract Ambiguity} and \textit{Extract Precondition}, and still achieved an 86.7\% approval. Among the transformations positively commented, we highlight \textit{``Following the logic that every action has a reaction, every test step must have an expected result [...],''} which reinforces the \textit{Fill Verification} transformation. Another comment received highlights that \textit{``In this case, some steps could be included in a `test setup' column, so that before the person/machine performs the steps, it ensures that it has the necessary resources to carry out the test.''} This comment reinforces transforming the \textit{Misplaced Precondition} smell.

Some respondents disagreed with some transformations offered and exemplified in our research. For example, in one of the tests presented in our survey, we used the following actions: ``\textit{Start Nautilus,}'' ``\textit{Maximize it's window,}'' and ``\textit{Start Firefox.}'' Our transformation considered three verifications for these three actions. However, a respondent mentioned the following: \textit{``[...] its not necessary a verification in this case.''} Another disagreement was in the \textit{Extract Verification}, where a respondent commented: \textit{``There's (verify xxxxx) two times. If you'll join to one block, don't repeat process.''} This case is connected to the provided example, where a test had two similar verifications written slightly differently but for the same action. This outcome highlights the need to identify semantic \textit{Test Clones}~\cite{hauptmann2013hunting} in future work.

\textbf{\textit{Answer to RQ$_{1}$.}} The online survey shows software testing professionals mostly agree with our proposals.

\subsection{Threats to Validity}

As a threat to \textbf{external validity}, collecting answers from only \NumberOfSurveyedProfessionals\ participants may bring bias in terms of generalization. We minimize this threat by relying on the experience of the participants to provide good responses. Also, we might experience a threat to \textbf{internal validity} when selecting the examples for the study. We minimize this threat by using real-practice examples from the Ubuntu OS. A last threat to the \textbf{conclusion validity} relates to all participants working for the same company. We minimize this selection bias with staff from different roles and hierarchical positions.
\section{A Tool to Remove Smells from Manual Tests}
\label{sec:tool}

We now present details on developing an NLP-based tool called Manual Test Alchemist. It extends prior work \cite{soares2023manual} on automatically detecting natural language test smells and implements their removal according to the transformations presented in Section~\ref{sec:catalog}. 

The Manual Test Alchemist tool development is centered on spaCy~\cite{spaCyWebsite}---a commercial open-source library for NLP---to implement our transformations for natural language test smells. The spaCy library features convolutional neural network models for part-of-speech (POS) tagging~\cite{pennTreebank}, dependency parsing~\cite{universalDependencies}, morphology parsing, and named entity recognition~\cite{spaCyWebsite}. Table~\ref{tab:tool:spacy} shows the analysis of the action sentence in \figurename~\ref{fig:catalog:remove ambiguities:example} provided by spaCy. Such analysis enables the implementation of our transformations.

For example, \textbf{adverbs of manner} can be identified through \texttt{POS=ADV}, \texttt{TAG=RB} and \texttt{Dependency=advmod} properties, \emph{i.e.,} the \textit{approximately} found in Table~\ref{tab:tool:spacy}. Such detection is used in the Extract Ambiguity transformation (Section~\ref{sec:catalog:catalog:removeAmbiguities}).

In another example, \textbf{imperative verbs} identify action steps \texttt{POS=VERB}, \texttt{VerbForm=Inf} and \texttt{Dependency=ROOT} properties, \emph{i.e.,} \textit{open} in Table~\ref{tab:tool:spacy}. This detection is used in the \textit{Separate Actions} (Section~\ref{sec:separate-actions-transformation}), \textit{Extract Action} (Section~\ref{sec:catalog:extract-action}), and \textit{Extract Precondition} (Section~\ref{sec:catalog:extract-precondition}) transformations.

\begin{table}[htb]
\caption{Example of spaCy's text analysis in detail.}
\label{tab:tool:spacy}
\footnotesize
\centering
\begin{tabular}{@{}lllll@{}}
\toprule
\textbf{Text} & \textbf{POS} & \textbf{TAG} & \textbf{Dependency} & \textbf{Morphology}       \\ \midrule
After         & ADP          & IN           & prep                &                           \\
approximately & ADV          & RB           & advmod              &                           \\
30            & NUM          & CD           & nummod              & NumType=Card              \\
seconds       & NOUN         & NNS          & pobj                & Number=Plur               \\
,             & PUNCT        & ,            & punct               & PunctType=Comm            \\
open          & VERB         & VB           & ROOT                & VerbForm=Inf              \\
the           & DET          & DT           & det                 & Definite=Def|PronType=Art \\
network       & NOUN         & NN           & compound            & Number=Sing               \\
manager       & NOUN         & NN           & dobj                & Number=Sing               \\
.             & PUNCT        & .            & punct               & PunctType=Peri            \\ \bottomrule
\end{tabular}
\end{table}

\begin{figure}[htb]
    \centering
    \includegraphics[width=\columnwidth]{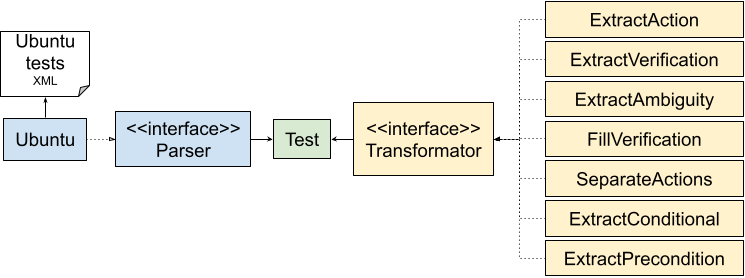}
    \caption{Simplified UML class diagram.}
    \label{fig:tool:architecture}
\end{figure}

Illustrated in Figure~\ref{fig:tool:architecture} is a simplified architecture of our tool, which functions by receiving an XML file of a test suite as its input. This test suite may contain multiple tests and is parsed using a customizable parser implementation into a consistent test format. Every implementation of the \texttt{Transformator} interface is responsible for identifying and adrressing a specific test smell.

The \textit{Extract Conditional} transformation results in the creation of new tests. Therefore, a relation between the execution order of transformations and their impact was identified. According to the implications shown in Section~\ref{sec:catalog}, the correction of some test smells may result in new ones. For instance, adjusting actions before \textit{Separate Actions} is important for proper functionality, and \textit{Fill Verification} should run last to address the \textit{Unverified Action} smell generated by \textit{Separate Actions} and \textit{Extract Action}. Hence, to minimize this effect, we defined the execution sequence to be the one presented on Table~\ref{tab:tool:transformations-smells}.

To provide a more detailed account of the tool’s inner workings, Algorithm~\ref{alg:tool:innerWorkings} presents the pseudocode of the \textit{Separate Actions} transformation. The algorithm highlights the key steps involved in identifying the presence of multiple action verbs within a single test step, separating the sentences related to each of these actions, allocating the separated sentences to new steps, and notifying the test professional that a verification must be written for the newly added step.

\begin{algorithm}
    \caption{The Separate Actions transformation (simplified)}
    \label{alg:tool:innerWorkings}
    \small
    \begin{algorithmic}
        \For {every test step}
            \State Search for action verbs;
            \If{action verbs $>$ 1}
                \For {every action verb}
                    \State Identify sentence connectors;
                    \State Separate sentences;
                    \For {every sentence}
                        \State Move sentence to new step;
                        \State Fill missing verification advise;
                    \EndFor
                \EndFor
            \EndIf
        \EndFor
    \end{algorithmic}    
\end{algorithm}
\section{Tool Evaluation}
\label{sec:tool-validation}

In this section, we present our tool evaluation. Here we aim to answer the research question \textbf{RQ$_{2}$:}\textit{``How precise is our tool in the task of removing natural language test smells?''}

\subsection{Planning}

The goal of this evaluation is to validate the tool in terms of Precision, Recall, and F-Measure metrics. To accomplish the goal, we first execute our tool against the \TotalNumberOfTestsWithinFiles\ natural language tests from Ubuntu OS, yielding  \TotalSmellOccurrences\ test smell occurrences. However, validating more than eight thousand transformations manually would be infeasible. Therefore, we perform a manual analysis in a sample of \SampleOfSmellOccurrences\ randomly selected smell occurrences (90\% confidence with 5\% margin of error, using the Cochran's Sample Size Formula~\cite{kotrlik2001organizational}). Table~\ref{tab:original and sampled} distributes the total and sampled occurrences per test smell.

\begin{table}[h]
\caption{Total and sample occurrences per test smell}
\label{tab:original and sampled}
\small
\begin{tabular}{@{}lrr@{}}
\toprule
\textbf{Test Smell} & \multicolumn{1}{l}{\textbf{Total}} & \multicolumn{1}{r}{\textbf{Sample}} \\ \midrule
Unverified Action      & 1,967 & 67 \\
Misplaced Precondition & 49   & 5  \\
Misplaced Action       & 345  & 8  \\
Misplaced Verification & 426  & 16 \\
Eager Action           & 2,663 & 69 \\
Ambiguous Test         & 2,656 & 90 \\
Conditional Test       & 279  & 9  \\
\midrule
\textbf{TOTAL}           & \TotalSmellOccurrences & \SampleOfSmellOccurrences \\\bottomrule
\end{tabular}%
\end{table}

For each transformation, three authors analyzed and validated whether the tool performed the transformations steps correctly or not, assuming that the transformation itself has been correctly chosen~\cite{soares2023manual}. In eight cases out of \SampleOfSmellOccurrences\ occurrences (0.03\%), they disagreed. However, the researchers reached a consensus in all cases after discussions. They collected the results in terms of true positives ($TP$)---the smell was corrected---, false positives ($FP$)---the smell was not corrected---, and false negatives ($FN$)---no correction was attempted.

\subsection{Results}
\label{sec:validation:results}

As results, we achieved the following: a Precision of \precision, a Recall of \recall, and a F-Measure of \fmeasure. Table~\ref{tab:tool-evaluation:percentages} illustrates our results for each transformation presented in Section~\ref{sec:catalog}.

Notice that the \textit{Fill Verification} transformation reached the highest $TP$ percentage. On the other hand, the \textit{Separate Action} transformation reached the lowest $TP$ percentage. Two extract transformations reached 25\% false positive rate: \textit{Extract Action} and \textit{Extract Verification}. Last but not least, the \textit{Separate Action} transformation reached 36.23\% false negative rate.

\begin{table}[h]
\caption{TP, FP, and FN per transformation}
\label{tab:tool-evaluation:percentages}
\small
\centering
\begin{tabular}{@{}lrrr@{}}
\toprule
\textbf{Transformation} & \multicolumn{1}{l}{\textbf{TP \%}} & \multicolumn{1}{l}{\textbf{FP \%}} & \multicolumn{1}{l}{\textbf{FN \%}} \\ \midrule
Fill Verification      & \cellcolor[HTML]{57BB8A}89.55\% & \cellcolor[HTML]{FDF0EF}2.99\%  & \cellcolor[HTML]{FFFFFF}7.46\%  \\ 
Extract Precondition & \cellcolor[HTML]{81CCA7}80.00\% & \cellcolor[HTML]{FFFFFF}0.00\%  & \cellcolor[HTML]{FFEEBD}20.00\% \\
Extract Action       & \cellcolor[HTML]{CDEBDC}62.50\% & \cellcolor[HTML]{E67C73}25.00\% & \cellcolor[HTML]{FFF8E5}12.50\% \\
Extract Verification & \cellcolor[HTML]{CDEBDC}62.50\% & \cellcolor[HTML]{E67C73}25.00\% & \cellcolor[HTML]{FFF8E5}12.50\% \\
Separate Actions          & \cellcolor[HTML]{FFFFFF}50.72\% & \cellcolor[HTML]{F2BBB6}13.04\% & \cellcolor[HTML]{FFD666}36.23\% \\
Extract Ambiguity         & \cellcolor[HTML]{8FD2B1}76.67\% & \cellcolor[HTML]{F3BFBB}12.22\% & \cellcolor[HTML]{FFFAEC}11.11\% \\
Extract Conditional & \cellcolor[HTML]{8AD0AE}77.78\% & \cellcolor[HTML]{F4C5C1}11.11\% & \cellcolor[HTML]{FFFAEC}11.11\% \\ \bottomrule
\end{tabular}%
\end{table}

\subsection{Discussion}
\label{sec:validation:discussion}

After having the results illustrated in Table~\ref{tab:tool-evaluation:percentages}, we performed an in-depth analysis on some of our \SampleOfSmellOccurrences\ sample tests to better understand potential problems of our tool.

When there is poor test writing (\emph{e.g.}, wrong phrases, excessive use of special characters, wrong formatting, \emph{etc.}), the tool malfunctions, mostly adding new special characters and breaking phrases into small ones. Also, in some tests, the special characters led the tool to apply the \textit{Extract Action} and \textit{Extract Verification} in a wrong way. For example, the tool detected the following action as a verification: ``\textit{Re-Check release-setting [...],}'' leading the \textit{Extract Verification} transformation to wrongly move the action. Here, spaCy separated ``\textit{Re-Check}'' in two words, and then the ``\textit{Check}'' verb was considered as a verification.

As our tool depends on semantic pattern matching coupled with syntax analysis, a transformation that relies on verb placement such as \emph{Separate Actions} is heavily affected by spaCy rules that do not comprehend some cases, such as in the following phrase: ``\textit{Select a plugin and configure it by doing click on `Configure'}''. In this example, the tool considered ``\textit{Configure}'' as an action. This way, it created a Step $S_i$  containing only the ``\textit{Configure}'' word.

To sum up, the results presented in Table~\ref{tab:tool-evaluation:percentages} provide a comprehensive overview of the effectiveness of each transformation, guiding further refinements and optimizations in our tool.

\textit{\textbf{Answer to RQ$_{2}$.}} Our analysis shows promising tool results, achieving a F-Measure of \fmeasure.

\subsection{Threats to Validity}
\label{sec:validation:threats}

As a threat to external validity, we have validated the tool in \SampleOfSmellOccurrences\ test smells occurrences, but all the sample is based only on one system (Ubuntu OS). This way, it may be difficult to generalize to other systems. To validate the tool, we performed a manual analysis, which poses threats to internal validity. The subjectivity of the process and the unawareness regarding technical items of the tests may led the authors who performed the manual analysis to disagreements and to commit errors. We minimize this threat by considering three persons to analyze the transformations. All disagreements have reached a consensus.
\section{Related Work}
\label{sec:related}

Hauptmann \emph{et al.}~\cite{hauptmann2013hunting} firstly applied the idea of smells to natural language tests. A set of seven test smells has been proposed and analyzed in real-practice tests. Ten years later, Soares \emph{et al.}~\cite{soares2023manual} complemented the list of natural language test smells with six more. Both works presented techniques to identify the smells. Whilst Hauptmann \emph{et al.} used keywords and metrics, Soares \emph{et al.} used newer NLP-based technologies. However, differently from our work, none of them presented neither transformations nor tools to remove the smells automatically.

Our work is also related to the software requirements field, since some test smells are similar to smells that appear in requirements. For example, Femmer \textit{et al.}~\cite{Femmer2017} define a set of requirements smells from ISO 29148. As an example, they define the \textit{Ambiguous Adverbs and Adjectives} requirement smell, which is quite related to the \textit{Ambiguous Test} smell. Also, Rajkovic and Enoiu~\cite{NALABS} proposed a tool, named NALABS, to identify requirements smells, like \textit{Vagueness}, \textit{Optionality}, and \textit{Subjectivity} (all of them also related to the \textit{Ambiguous Test}). Recently, Fischbach \emph{et al.}~\cite{jannik-jss} implemented a tool to generate a minimal set of test cases to a given set of requirements containing conditionals. The requirement smell explored is very related to the \textit{Conditional Test}. These works provide tools to identify the requirements smells, but not to remove them automatically as we do in the context of natural language test smells.

Fischbach \emph{et al.}~\cite{Fischbach-PROFES-2021} conducted a survey to understand how requirements engineers interpret conditionals in software requirements. The authors found that the conditionals led the engineers to interpret the requirements in an ambiguous way. In our work, we considered an analogous smell, \textit{i.e.}, the \textit{Conditional Test} smell. We also conducted a survey and our results show that the majority of the respondents agrees with our transformation to remove the \textit{Conditional Test} smell.
\section{Implications for Practice}

As implications for practitioners, our catalog and tool might provide a standardized and systematic approach to address natural language test smells, promoting consistency in test practices across projects and teams and ensuring that similar issues are addressed uniformly. Also, they can help to improve test quality and maintenance. In addition, our tool avoids repetitive tasks related to identifying and fixing test smells, which provides time and cost savings and could lead practitioners to focus on more complex and valuable aspects of testing. Last but not least, our research could also bring implications for the training and development of software testing professionals. By providing a catalog of transformations along with a tool to remove test smells, we introduce learning resources for understanding what these smells are and how to address them.

For researchers, our catalog represents a step forward towards more research on NLP-based techniques to avoid and remove smells in natural language tests. For instance, researchers can build upon our catalog as a starting point to develop more transformations and improve our propositions. Our work might also be a preliminary benchmark for evaluating and comparing different NLP-based techniques and tools to remove natural language test smells.
\section{Concluding Remarks}
\label{sec:conclusion}

In this paper, we introduced a catalog of transformations to remove seven natural language test smells. We assessed the quality of our catalog by recruiting \NumberOfSurveyedProfessionals\ software testing professionals. The professionals found the transformations valuable, which is supported by the high average acceptance rate. We also introduced a tool that implements the catalog. We executed the tool in \SampleOfSmellOccurrences\ occurrences of test smells and achieved \fmeasure\ of F-Measure rate.

As future work, we intend to (i) increase the set of transformations to include other smells (\textit{e.g.}, \textit{Tacit Knowledge}~\cite{soares2023manual}, \textit{Test Clones}~\cite{hauptmann2013hunting}, and \textit{Long Test Steps}~\cite{hauptmann2013hunting}); and (ii) use Large Language Models (LLMs) to improve the effectiveness of our tool.

\begin{acks}
This work was partially funded by CNPq 312195/2021-4, 310313/2022-8, 403361/2023-0, 443393/2023-0, 404825/2023-0, 315840/2023-4, FAPEAL 60030.0000000462/2020, 60030.0000000161/2022, and FAPESB PIE002/2022 grants.
\end{acks}

\bibliographystyle{ACM-Reference-Format}
\bibliography{bibliography}

\end{document}